\documentstyle[seceq]{ptptex}

\notypesetlogo  
\title{
 Supergravity, AdS/CFT Correspondence,  
and Matrix Models
}

\author{
Tamiaki {\sc Yoneya}\footnote{  E-mail address: tam@hep1.c.u-tokyo.ac.jp}
}

\inst{
Institute of Physics,  University of Tokyo 
\\
 Komaba, Meguro-ku, Tokyo 153-8902
}

\abst{
The recent developments towards the possible 
non-perturbative formulation of string/M theory 
using supersymmetric Yang-Mills 
matrix models (SYMs) are discussed.  
In the first part, we give a critical review 
on the status of our present understanding, 
focusing on the connection of the D0-brane matrix models to supergravity 
and its relevance to the 
so-called Matrix-theory conjecture. We also discuss  
some problems concerning
 the conjectured relation between supergravity in 
AdS background and SYM from the viewpoint 
of D-brane interactions.   We present a qualitative 
argument showing how the boundary condition 
at AdS boundary dictates the correlators on the 
large $N$ system of source D-branes.   
Then, in the final part, we 
turn to the question how to formulate the condensation of graviton in matrix models, taking the 
simplest example of 
type IIB matrix model.  We 
argue the emergence of a
 hidden symmetry GL(10, R), beyond the 
manifest Lorentz symmetry  SO(9,1), 
by embedding U($N$) model into 
 models with higher $N$ and by treating the whole 
recursive series of models simultaneously. This suggests 
a possible approach toward background independent 
formulations of matrix models. }

\begin{document}
\newcommand{\EQ}{\begin{equation}}
\newcommand{\EN}{\end{equation}}
\newcommand{\EQAN}{\begin{eqnarray*}}
\newcommand{\EQNN}{\end{eqnarray*}}
\newcommand{\EQA}{\begin{eqnarray}}
\newcommand{\EQN}{\end{eqnarray}}
\newcommand{\e}{{\rm e}}
\newcommand{\Sp}{{\rm Sp}}
\newcommand{\Tr}{{\rm Tr}}
\newcommand{\p}{\partial}
\newcommand{\EQAnn}{\begin{eqnarray*}}
\newcommand{\EQNnn}{\end{eqnarray*}}

\maketitle

\section{Introduction}
One of the most remarkable insights gained through the 
recent developments in the studies of duality 
symmetries of string theory is the possibility 
of formulating nonperturbative string theory 
in terms of supersymmetric Yang-Mills theories.  
  From an ordinary 
viewpoint of perturbative string theory, 
Yang-Mills theories  are 
regarded as the low-energy effective theories  
for describing interactions of gauge-field 
excitations of strings.  
The discovery of crucial roles \cite{pol} 
played by Dirichlet branes 
(D-branes) 
for realizing string dualities, however,  paved a way 
toward possible reformulations of string theory 
using new degrees of freedom other than the 
fundamental strings, on the basis of an entirely 
different interpretation of Yang-Mills fields.   
Since  string field 
theories assuming the 
fundamental strings themselves to be the basic 
degrees of freedom do not seem 
to be appropriate for nonperturbative 
studies of the theory, such an alternative possibility has long been sought, but has never been materialized in concrete 
form.  

In the present report, I would like to review the 
status of  Yang-Mills matrix models 
from the viewpoint of asking the question,  
``{\it Why could the models be the 
theory of quantum gravity?}"  
In this written version of the talk, some of the 
subjects which I have presented in the YITP workshop,  
held in succession to the Nishinomiya Yukawa symposium, 
will also be included and expanded.   

We will start from the so-called Matrix theory which 
was proposed first and has been a focus of most 
intensive studies.  Next we will turn to the 
 so-called AdS/CFT(SYM) correspondence. 
The purpose of this first part is 
to review the known results critically and provide 
a few new observations on some 
unsolved issues.   
We will then proceed to the issue why 
Yang-Mills matrix theories could be the 
models for quantum gravity. A special emphasis 
will be put on possible hidden symmetry structure 
which would ensure the emergence 
of general covariance at long distance regime.  
The last part discussing the problem of background 
independence is a 
preliminary report from a still unfinished project and 
will be of very speculative character.  

\section{Yang-Mills matrix models and supergravity}
The D-branes are  objects carrying 
Ramond-Ramond (RR) charges. They are necessary for 
realizing duality symmetry among various 
perturbative vacua of string theory, since the 
transformations associated with duality 
interchange Neveu-Schwarz-Neveu-Schwarz 
(NS-NS) and RR charges. 
In the case of perturbative closed-string theories, 
it is known that the type IIA or IIB theory 
allows even or odd (spatial) dimensional D-branes, 
respectively.  
In particular, the lowest dimensional objects are 
D0-brane (D-particle) in IIA and D(-1)-brane (D-instanton) 
in IIB.  

In low-energy effective field 
theory, namely, IIA or IIB supergravity, D-branes are 
represented as solitonic classical solutions. 
From the viewpoint of ordinary world-sheet formulation 
of the theories, they are described as collective 
modes of fundamental strings, in which the collective 
coordinates can be identified with the 
space-time coordinates at the 
boundaries of open strings with Dirichlet condition. 
In old perturbative string theory, it has been thought 
that open strings cannot be coupled 
to closed strings consistently, since they necessarily
break the N=2 supersymmetry of closed-string 
sector.   In our new understanding,  
the partial breakdown of supersymmetry 
just indicates the existence of 
D-branes as physical objects, and 
the remaining supersymmetry is reinterpreted as 
the manifestation of 
the BPS property of D-branes.  

To describe the dynamics of D-branes, we have to 
therefore study coupled 
systems of closed strings and open strings 
 with dynamical Dirichlet boundary 
conditions.  
Here it is worthwhile to recollect an old but well-known 
formulation of open strings, namely, Witten's 
string field theory \cite{wittensft}.  The latter only uses open-string 
fields as the dynamical degrees of freedom.  
Nevertheless, it includes the whole dynamics 
of interacting closed-open strings.   Namely, 
it is possible to describe the dynamics of closed 
strings in terms of open-string degrees of 
freedom without explicitly introducing fields corresponding 
to closed strings.  
This remarkable property is 
actually a consequence of the old $s$-$t$ duality 
which is the basis for conformal invariance 
of the world-sheet string dynamics.  
Furthermore, if there were circumstances 
where we can neglect the  excitation modes of 
open strings higher than the lowest Yang-Mills degrees of 
freedom,  we can even imagine 
 situations where the whole dynamics  including quantum 
gravity effect can be described by Yang-Mills 
theories.    Let us  briefly review some representative 
proposals along this line. 

\subsection{D-particle model or Matrix theory} 
The first such model is called `Matrix theory'. 
The model is based on the 1+0-dimensional 
Yang-Mills theory with maximum (N=16) supersymmetry, 
which is obtained by the dimensional reduction 
from 10 (=1+9) dimensional N=1 supersymmetric 
Yang-Mills theory.  
\EQA
S \hspace{-0.1cm}&=&\hspace{-0.2cm} \int dt \, \Tr
\Bigl( {1\over 2g_s\ell_s} D_t X_i D_t X_i + i \theta D_t \theta
+{1 \over 4g_s\ell_s^5} [X_i, X_j]^2 -
{1\over \ell_s^2}\theta \gamma_i [\theta, X_i]\Bigr) \\
\hspace{-0.2cm} &=&\hspace{-0.1cm} \int dt \, 
\Tr 
\Bigl( {1\over 2R} D_t X_i D_t X_i + i \theta D_t \theta 
+{R \over 4\ell_P^6 } [X_i, X_j]^2 -
{R\over \ell_P^3}\theta \gamma_i [\theta, X_i]\bigr) \, ,
\label{d0matrixmodel}
\EQN
where $\ell_s$ and $ g_s$ are string length and coupling constants, 
respectively. In the second line, we have introduced the 
11 dimensional parameters of M-theory, 
compactification radius $R=g_s\ell_s$ along the 
11th direction (10th spatial direction) and the 
11 dimensional Planck length $\ell_P= g_s^{1/3}\ell_s$.  
The Higgs field $X_i \, (i=1, 2, \ldots, 9)$ are 
dimensionally reduced U($N$) gauge-field matrices whose diagonal 
components are identified with the collective 
coordinates of 
$N$ D-particles, while the 
off-diagonal components are 
the fields of lowest open string modes 
connecting the D-particles.  The 16 component Grassmann 
(hermitian) matrices $\Psi$ transforming 
as SO(9) spinor are the super partner of the 
Higgs fields. 

In the matrix theory conjecture 
proposed in ref. \cite{bfss}, 
this action is interpreted as the effective action 
of the theory in the Infinite-Momentum Frame (IMF) 
where the 11th total momentum $P_{11}$ is 
taken to be infinitely large.  Following the M-theory identification 
of the 11th momentum  with the RR 1-form charges of 
D-particle,  it is assumed that 
\EQ
P_{11}= N/R\, .
\EN
Thus, for any finite fixed $R$, the IMF limit corresponds to 
taking the large $N$ limit, $N\rightarrow \infty$.  
In other words, this requires that the IMF Hamiltonian 
$P^-=P^0-P^{11}$ must have nontrivial dynamics 
in the part which scales as $1/N$. 
One of the  reasonings  for this conjecture is that 
in the IMF frame (the part of) the (super) Poincar\a{`}e symmetry 
is reduced to (super) Galilean symmetry 
in the 9 dimensional transverse space and the above 
action precisely exhibits that symmetry.  
In particular,  to be a theory in 11 dimensional 
space-time, it should exhibit the $N=1$ supersymmetry 
in 11 dimensions which amounts to  $N=2$ supersymmetry 
in 1+9 dimensions.  Indeed the model has, 
 in addition to 
the $N=16$ supersymmetry in 1+0 dimensions inherited from 
10 dimensional $N=1$ supersymmetry
\EQ
\delta^{(1)} \theta = 
{1\over 2} ({1\over R}D_tX_i\gamma^i 
+ {i\over 2\ell_P^3}[X_i, X_j]\gamma^{ij})\epsilon^{(1)}, 
\EN
a trivial supersymmetry under 
\EQ
\delta^{(2)} \theta = \epsilon^{(2)} , 
\EN    
where $\epsilon^{(1)}$ and $\epsilon^{(2)}$ are 
two independent constant Majorana spinors.  
The algebra of these two supersymmetry transformations 
closes with central charges up to a 
field-dependent gauge transformation. 
For a single D-particle state at rest 
as the simplest example, the first 
symmetry is unbroken, 
corresponding to the 
BPS property of the state, while the 
second is broken.   
Furthermore, the dimension of the 
multiplet of single particle states 
fits to the desired multiplet corresponding to the 
first Kaluza-Klein mode of the 11 dimensional 
multiplet containing massless graviton and 
gravitino : The 16 component Grassmann coordinate 
$\theta$ leads to $2^{16/2}=256=128+128$ dimensional representation 
of transverse SO(9) ($\sim$ Spin(9)) group, which 
is precisely the physical dimension of the 11 dimensional 
supergravity multiplet.  

Of course, the 
Galilean symmetry is not sufficient to justify the 
decoupling of the higher string modes.  That is the 
crucial dynamical assumption of the model.  
A piece of  evidence for this conjecture comes from the 
old observation made long time ago in connection 
with the theory of membrane. Namely, 
the same model can be interpreted as a special 
regularized version of the membrane action 
\cite{dhn} 
in the light-cone gauge in 11 dimensions.  
The large $N$ limit in this interpretation is 
nothing but the continuum limit. 
The fundamental string of 10 dimensional 
type IIA theory is identified with the membrane 
which is wrapped along the compactified 
11th direction.  If this really works, 
it is quite natural to expect that 
after taking the appropriate large $N$ limit 
the model would reproduce the whole dynamics 
including the effect corresponding to 
higher string modes.  
The crucial new observation here 
is that the model should be interpreted as describing 
the arbitrary multi-body systems of membranes or 
D-particles.  This solves the long-standing problem 
in the formulations of supersymmetric membranes, 
namely, the difficulty of continuous energy 
spectrum.  The energy spectrum of the system 
{\sl must} be continuous from zero, to be the theory of 
multi-body system including the massless 
particles.  For the validity of this interpretation, 
it is necessary that there exists one and only one 
threshold bound state which is 
identified with the single-particle 
graviton supermultiplet for each fixed 
$N$ of U($N$).  At least for $N=2$, this is 
consistent with the Witten index of the model \cite{index}. 

Another impetus for this model is the proposal that 
the model might be meaningful even for finite $N$. 
That is, Susskind \cite{suss} suggested that the model for finite $N$ should be interpreted following the 
framework of the so-called discrete light-cone  
quantization (DLCQ), in which the compactification is 
made along the light-like direction $x^-=
x^{11}-x^0$ instead of the space-like 
direction.  Such a formalism has often been discussed 
to regularize gauge field theories. 
In fact, this proposal can be
 related to the IMF interpretation by 
considering the limit of small $R$ keeping $N$ fixed, 
which is another way of making $P_{11}$ large. 
If we boost the system simultaneously with taking 
this limit, we can keep the longitudinal momentum 
$P_-$ finite and the condition of  compactification 
is imposed on the $x^-$ direction with finite 
compactification radius in the small $R$ limit.    
This is essentially the argument 
given in ref. \cite{seisen}.  Equivalently, using the 
original frame with sufficiently small $R$, the 
compactification condition $x^{\pm}\sim x^{\pm}+2\pi R$ 
in the space-lime 11th dimension 
can be approximated by the light-like 
condition  $(x^{-}, x^{+})\sim (x^{-}+2\pi R, x^{+})$,  
since in the limit we are only interested in the small 
longitudinal energy $P_+\sim P_i^2/2P_-$ proportional to 
$R$  while $P_-\sim P_{11} \sim O(1/R)$ becomes 
large. 

Now since the limit forces the 11 dimensional Planck length 
small compared to the string length, $\ell_P \ll \ell_s$,  
we expect that the interaction of D-particles can be 
described by lowest open string modes at least at 
distance scales shorter than the string length, 
according to the result of ref. \cite{dkps}.  
Also, the 11 dimensional Newton 
constant $G_{11} \sim g_s^3 \ell_s^9$ 
becomes small in this limit.  Thus from the viewpoint 
of closed strings or membranes, the interaction 
of D-particles should be approximated well 
by classical supergravity at least for distance 
scales much larger than the string length $\ell_s$.  
If one naively {\bf assumes} that Matrix theory is correct and 
that the justification 
of Matrix theory comes {\bf solely} from the infinite 
momentum limit,  one might expect that 
Matrix theory for sufficiently small $R$ with 
finite $N$ must reproduce the classical 
supergravity at all distance scales which are larger than 
the 11 dimensional Planck length.  This would in particular 
require that the lowest open string mode alone  
describes correctly the gravitational 
interaction of D-particles for such wide 
ranges of distance scales, namely, 
 from the infinite large distances 
all the way down to near the 11D Planck length which is 
far  below the string scale.  
This would be quite a surprising conclusion, 
since we usually think that the duality between 
open and closed strings is due to the existence 
of full tower of higher string modes on both 
sides of closed and open strings.   In particular, 
the effective dynamics near the string length 
after eliminating the higher-string modes would 
necessarily be non-local either in terms of the 
lowest graviton fields alone or of the lowest gauge modes alone .  
Before discussing further the meaning 
of this and where this 
naive expectation may be invalidated, 
let us briefly review known results related to this issue.

\subsection{Matrix theory {\it vs.} supergravity}
     In fact, as discussed in ref. \cite{dkps}, 
supersymmetry 
ensures that the above conclusion is 
indeed true at least in the one-loop approximation 
in terms of open string computation.   
The two-body interaction, $v^4/r^7$,  of D-particles 
in the leading approximation with respect to the 
expansion in velocity is correctly reproduced 
by only the lowest open string modes.  
Namely, the same expression is valid for the 
large $r$ region where the only lowest modes of the 
closed string couple, as described by supergravity. 
Furthermore, at least for two-body interactions, 
 a non-renormalization theorem \cite{pss}  is established 
demanding  that the one-loop result for the leading 
term is not 
renormalized by higher order effects.   
This theorem can be generalized to the next order 
$v^6$ for the two-body interaction and 
is consistent with the result of explicit two-loop 
computation \cite{bbpt} of the two-body interactions.  

Whether similar non-renormalization theorem 
is valid for more general multi-body interactions 
is not known.  Extension of the 
argument given in \cite{pss} to general $N$-
body interactions \cite{lowe} 
is difficult.  In general, however, we hope that 
some symmetry together with certain 
additional inputs would fix the theory of 
gravity completely.  For example, we believe that 
general covariance and locality uniquely 
lead to General Relativity at sufficiently large distances.  
So the question is whether the supersymmetry of the 
matrix model (\ref{d0matrixmodel}) is sufficient 
to ensure the general coordinate invariance 
at large distances 
as interpreted in 11 dimensional space-time. 
If we assume the existence of  massless 
graviton supermultiplet and Lorentz 
symmetry in the flat background in 11 dimensions, 
only consistent low-energy effective theory is believed to 
be supergravity.  Establishing Lorentz invariance 
of the model in the limit $R, N \rightarrow \infty$ 
would thus be most desirable.  
At least for membrane approximation, this is 
very plausible.  However, the 
membrane approximation is not sufficient 
to establish the Lorentz symmetry, since 
the interpretation of the matrix model is really 
very different from membrane as emphasized already. 
Unfortunately no concrete proposal 
for general case has been given.  It is thus desirable to 
perform explicit computations for multi-body 
interactions.  

 Let us here briefly review the result of explicit 
 computations of 3-body interaction of D-particles at finite $N$.  
A scaling argument shows that the 
effective lagrangian of the 3-body interaction 
must take the following form 
\EQ
L_3 \sim {G_{11}^2\over R^5}{v^6\over r^{14}}\,  ,
\EN
where the factor $v^6/r^{14}$ only indicates 
power behaviors with respect to relative velocities ($v$) and 
to relative distances ($r$).  The power $R^{-5}$ 
with respect to 
the compactification radius is required 
by boost invariance along the 11th direction. 
Note that in terms of the Yang-Mills coupling 
$g_{{\rm YM}}^2 \propto g_s$ 
the factor $G_{11}^2/R^5\propto g^2_{{\rm YM}}$ 
corresponds to the two-loop contribution.  
For small $g_s$, the compactification radius is 
small, but the Newton constant  is also vanishing 
 such that the 
expansion parameter $G_{11}^2/R^5$ is arbitrarily small.  
This indicates that the regions of validity of classical 
supergravity and perturbative computation in 
the matrix model might overlap, if only the parameters 
are concerned neglecting the real roles of the 
dynamical variables. Therefore it is not unreasonable 
if the matrix model with finite $N$ is able to reproduce 
supergravity results to some finite orders with 
respect to the Newton constant.      

In classical supergravity,  we can derive the following 
explicit form for the interaction lagrangian 
\EQ
L_3 = L_V +L_Y.
\EN
\EQ
L_V=  -\sum_{a,b,c}{(15)^2 N_aN_bN_c\over 64 R^5M^{18}}
v_{ab}^2 v_{ca}^2 (v_{ca}\cdot v_{ab})
{1\over r_{ab}^7}{1\over  r_{ca}^7} .
\label{eq256}
\EN
 \EQA
L_Y  &=&-\sum_{a, b, c} 
{(15)^3 N_aN_bN_c \over 96(2\pi)^4R^5M^{18}} 
\Bigl[
-v_{bc}^2v_{ca}^2(v_{cb}\cdot \nabla_c)
(v_{ca}\cdot \nabla_c)\nonumber \\
&&+{1\over 2} v_{ca}^4(v_{cb}\cdot \nabla_c)^2  
+{1\over 2}v_{b c}^4 (v_{ca}\cdot \nabla_c)^2
\nonumber \\
&&-{1\over 2}v_{ba}^2
v_{ac}^2(v_{cb}\cdot \nabla_c)(v_{bc}\cdot \nabla_b)
 \nonumber\\
&&+{1\over 4}v_{bc}^4 
( v_{ba}\cdot \nabla_b)(v_{ca}\cdot \nabla_c)
\Bigr]\Delta(a,b,c) 
\label{eq258}
\EQN
where
\[
\Delta(a,b,c) \equiv
\int d^9 y {1\over |x_a-y|^7 |x_b -y|^7|x_c-y|^7} 
\]
\EQ
= {64(2\pi)^3 \over (15)^3}
\int_0^{\infty} d^3\sigma 
(\sigma_1\sigma_2 + \sigma_2\sigma_3 + \sigma_3\sigma_1)^{3/2}
 \exp \bigl( 
-\sigma_1|x_a-x_b|^2 -\sigma_2|x_b -x_c|^2 -\sigma_3|x_c-x_a|^2 \bigr) 
\label{eq247}
\EN
and the indices $a, b, c, \ldots$ label the D-particles 
whose masses are $N_a/R, N_b/R, N_c/R, \ldots$. 
The Planck mass $M=1/\ell_P$ is defined by 
$G_{11}=2\pi^5/M^9$.  
The above separation into V-part and Y-part 
roughly corresponds to the contributions from 
the seagull-type 
diagrams and the diagrams with one 3-point self-interaction 
of graviton, respectively.  Because of the BPS property, 
the contribution from the Y-part  vanishes 
whenever any two D-particles have parallel velocities.  

On the side of Matrix theory,  we compute the 
scattering phase shift in the eikonal approximation. 
Each of the D-particles 
with masses $N_a/R, N_b/R, N_c/R, \ldots$ 
is approximated as a cluster of corresponding number 
($N_a, N_b, N_c, \ldots$) 
of D-particles, moving parallel within each 
cluster, with the smallest unit of 
mass $1/R$.  
We can again separate the two-loop contributions 
into V-and Y- types. The Y type contribution 
only comes from the diagrams with two 3-point 
vertices.  The V-type contribution comes from the 
diagrams with one 4-point vertex and also 
from the diagrams with two 3-point vertices 
in which one of the propagators is canceled 
by the derivatives acting on the 3-point vertices.  
For more details, the reader should consult our original 
papers \cite{oy1}\cite{oy2}.  It turns out that 
the V-type contribution to the eikonal phase shift 
can be written as the time integral of the above 
lagrangian $L_V$.   For the Y-type contribution 
which is vastly more complicated,  
we have confirmed that the result of explicit time integration 
of the lagrangian $L_Y$ precisely agrees with the 
phase shift obtained from the matrix model.  
We can also show the precise correspondence at the level of 
  the equations of motion on both sides including the effect of recoil \cite{oy2} to the present order of approximation.  

In the absence of general arguments which may guarantee 
the agreement between matrix theory with finite $N$ and supergravity 
in the long-distance limit, the above 3-body computation is 
the strongest evidence so far for the validity of 
Matrix theory conjecture in its DLCQ 
interpretation.  A related 
computation involving the 
nonlinear graviton interaction 
has also been done for graviton scattering in an orientifold 
background \cite{daniel} and exact agreement is verified. 
Extension of these 
computations to higher-loop/body interactions 
is not difficult at least conceptually, 
but is technically formidable and 
no complete computations for higher cases have 
been reported yet, except for a partial computation 
\cite{dine} which indicates some signal 
for a possible discrepancy with supergravity at 3-loop order.  

It should be emphasized here that 
given only the connection 
between the matrix model as the low-energy effective 
theory for D-particles in type IIA theory, on one hand,  and 
the connection of supergravity and closed strings on the 
other hand, 
the agreement of D-particle scatterings 
between supergravity and matrix model at 
arbitrary {\bf large} distances is in no sense a logical 
consequence.  Remember that 
the argument of ref. \cite{seisen} 
is not applicable at distances near and larger than the 
string scale.    
Suppose that the original BFSS conjecture 
that in the large $N$ limit (and for fixed 
$R$) the agreement 
is achieved is true.  Then the disagreement 
between supergravity and matrix theory at 
finite $N$ with small $R$, if it indeed occurs, 
 must be due to the neglect of 
bound-state effect in forming the states of 
D-particles with large longitudinal momentum. 
Namely, no matter how $R$ is small, 
only for sufficiently large $N$ we would 
expect that the effect of higher string modes which 
would ensure the validity of the $s$-$t$ duality between 
open and closed strings is correctly taken into account.  
To check whether matrix theory can give 
sensible results in this way is, however,  a very difficult 
problem, since for large $N$ the size  of 
graviton is known 
to grow indefinitely \cite{bfss} in the limit, and hence 
we have to deal with complicated many body dynamics 
of D-particles (or partons).  

If that is the case, it is desirable to have definite criteria 
on the basis of which we can assess various 
situations such that agreements or disagreements 
between supergravity and matrix theory  can 
be predicted by general arguments.   
For example, if we assume the correspondence between 
supgergravity and D0-matrix model following 
Maldacena's general conjecture \cite{maldacena} 
which will be the subject of  the 
next subsection,  the validity of the classical 10 
dimensional supergravity 
description is expected  for the distance scales\cite{itzak}
$\ell_PN^{1/7} \ll r \ll \ell_P N^{1/3}$ , 
where the first and the second inequalities 
come from the weak coupling condition and the small 
curvature condition, respectively.  At the lower end, 
the effective radius along the 11th direction becomes of the 
same order as $\ell_P$. Thus from the viewpoint of 
11 dimensions, it should not be regarded as the limit 
of the supergravity description as long as the 
curvature radius is much larger than the Planck length, 
although 10 dimensional description is no more 
valid. However the upper limit  indicates that the agreement with classical 
supergravity at  arbitrarily large distances can only 
be achieved in an appropriate large $N$ limit.  
Namely, the parameter $N$ plays effectively a role of 
infrared cutoff for the theory, not only with respect to the 
11th direction but also to the transverse space in the bulk,  
as we have argued before from the correspondence between the 
matrix model and a regularized theory of membranes.  
In other words, the low-energy long distance physics 
of supergravity is governed by the high-energy 
physics of open strings where in general 
we cannot neglect higher string (or membrane) modes. 

However, we should also keep in mind that the near-horizon limit,  
approximating the factor $1+ q/r^7$ by $q/r^7$,  is only valid 
if $r\ll (g_sN)^{1/7}$ where $q\propto g_sN$ , and hence the argument 
cannot be extended to the upper limit $\ell_P N^{1/3}$ 
in the strong coupling region. Thus strictly speaking 
we can not be sure whether Matrix theory reproduces the 
large distance behavior of classical supergravity in the 
large $N$ limit for {\bf fixed} $g_s$ when $g_sN\gg 1$, 
even if we assume the validity of the Maldacena 
correspondence in its origianl form. 
Of course, the Maldacena 
conjecture only proposes  sufficient conditions, 
and hence does not necessarily 
exclude the possibility that the region of validity 
extends beyond these conditions because of 
some (hidden) symmetry constraints depending on the type of 
physical quantities in question.  
We should also expect from a more 
general viewpoint that some 
(but perhaps already `built-in') symmetry must be responsible for the matrix model to reproduce supergravity 
in spite of the rapid growth of the size of graviton 
in the large $N$ limit.  Perhaps the precise 
agreement of 3-body interaction in the above finite $N$ 
calculation should be interpreted as a partial 
indication for the existence of such higher symmetry. 

Concerning the question on why the matrix model 
can be the theory of gravity, one of the other crucial 
unsolved problems is how to extend the model to 
general curved backgrounds.  
It has been argued that any simple modification of the 
quantum mechanical lagrangian (\ref{d0matrixmodel}) 
for the 
curved space cannot reproduce the 
supergravity result even at the order $v^4$ 
for the D-particle interaction in curved space for finite $N$. 
This seems to indicate that the curved background 
cannot, in general,  be described by finite $N$ models.    
Indeed, this is not unreasonable since to really modify 
the background in a self-consistent fashion within the 
framework of M-theory, we must consider the 
condensation of gravitons.  It is difficult to treat finite 
condensation of graviton in the present framework 
of Matrix theory which assumes fixed $N$ however 
it is large.  In the last part of 
this talk, I will give a preliminary consideration 
on the graviton condensation 
in a simpler case of type IIB matrix model \cite{ikkt}. 
For the possibility of modifying the action 
to curved backgrounds, an axiomatic 
approach called D-geometry \cite{douglas} has been 
suggested.  We have to await to see whether this approach 
can resolve the above issues.  We also mention 
a recent important 
work \cite{taylor} discussing the change of the 
background in Matrix theory, on the basis of 
one-loop computations of the interactions 
between an arbitrary pair of extended objects in 
the theory.  For an earlier approach from the 
viewpoint of membrane dynamics in curved background, 
see \cite{dewit} and references therein. 

\subsection{AdS/SYM correspondence} 
As already mentioned, another recent development closely related to  Matrix theory 
 is the conjectured correspondence \cite{maldacena} 
between supergravity in anti de Sitter background on one hand and super Yang-Mills theory of D-branes on the other. 
This is essentially based on the following two observations.  
Firstly, the low-energy (low-velocity) dynamics of many 
D-branes which are situated almost on top each 
other is well described by the effective 
super Yang-Mills theory for any finite $N$, 
since we can assume the decoupling of higher 
modes of open strings for the same reason as 
we have argued in the case of Matrix theory. 
Secondly, the field-theory description 
in terms of supergravity is expected to be simultaneously 
effective when the curvature near the horizon 
becomes sufficiently small compared with the 
string length.  In the case of D3-brane, in particular, 
the curvature radius at the horizon is of order 
\EQ
R_c \propto (g_{{\rm YM}}^2 N)^{1/4}\ell_s  . 
\EN
D3-brane is special in that the background dilaton 
is constant, and thus can be made arbitrarily small 
by keeping $R_c$ large ($g_{{\rm YM}}^2 N\gg 1$) 
if $N$ is sufficiently large. 
Then by the duality between open and close strings, 
we naturally expect that the descriptions of the 
dynamics of D3-branes 
in terms of supersymmetric Yang-Mills theory or 
type IIB supergravity are both valid.  In other words, 
super Yang-Mills theory in the large $N$ limit is 
expected to  be 
`dual' to supergravity with the D3-background in the 
near horizon limit.  The D3-brane metric 
in the near horizon limit is the direct product, 
AdS$_5\times$S$^5$,  of the five 
dimensional anti de Sitter space-time AdS$_5$ and 
five dimensional sphere S$^5$.  Thus the metric has 
isometric symmetry under the group 
$SO(4,2)\times SO(6)$.  Correspondingly, 
Yang-Mills theory 
for D3 branes are the $N=4$ superconformal 
Yang-Mills theory which has the same conformal symmetry 
$SO(4,2)$ and global  R-symmetry $SO(6)$.   
Using this correspondence, we can for example 
predict the spectrum of the superconformal Yang-Mills 
theory in the large $N$ limit by analyzing the 
Kaluza-Klein spectrum around the AdS$_5 \times $S$^5$.  

A more concrete prescription which allows us to 
connect correlators of both sides has been proposed 
in \cite{gkp}.  It essentially says that the 
effective action of supergravity for the supergravity fields 
which satisfy appropriate boundary condition at the 
boundary of the AdS space (opposite to the 
horizon) is the generating functional 
for the correlators of super Yang-Mills theory.  
The external fields for the latter 
generating functional coupled to operators of 
the Yang-Mills theory are nothing but the boundary 
value of the bulk fields in supergravity.   
Many computations of correlators have been done 
based on this conjecture.  However, it seems  that this prescription has 
never been derived logically from the 
duality between open and closed strings. 
For example, it is not clear why the boundary condition  
at the boundary of the AdS space-time can dictate the choice of 
operators of the large $N$ Yang-Mills theory, 
since naively the D-branes as the heavy source 
producing the AdS background seem to be situated at the {\sl opposite}  
`boundary' of the AdS space. 
In the following,  
we will first discuss some interesting aspect related to the 
correspondence of conformal symmetries on both 
sides, and then 
come back again to the issue of correlators later.

The metric of the AdS$_5\times$S$^5$ is given by
\EQ
ds^2 = \alpha'\Bigl(
{R_c^2 \over U^2}(dU^2
+ U^2d\Omega_5^2) + {U^2\over R_c^2}dx^2_{4} 
\Bigr)  , 
\label{ads5metric}
\EN
where $U=r/\alpha' \, \, (\alpha'\propto \ell_s^2)$ 
is the energy of an open string stretched from the 
source D3-branes at the origin to a probe D3-brane.  
The four dimensional flat metric $dx_4^2$ 
is interpreted as describing  the world-volume of the source 
consisting of $N$ 
D3-branes which are almost coincident to each other. 
The special conformal transformation in 
the SO(4,2) isometry is 
\begin{eqnarray}
\delta_K x^a &=&  -2\epsilon\cdot x\, x^a +
\epsilon^a x^2 +\epsilon^a {R_c^4\over U^2}  ,
\label{specialadsx}\\ 
\delta_K\,  U &=& 2\epsilon\cdot x \, U .
\label{specialadsu}
\end{eqnarray}
As noted originally in ref. \cite{maldacena},  the 
existence of the last term $R_c^4/ U^2$ leads to a nonlinear 
and field-dependent 
transformation for the dynamical coordinates of 
the probe D3-brane.  The latter property constrains the action to be the Dirac-Born-Infeld action for the probe D3-brane 
in the background of the source D3-branes, 
with a help of a supersymmetric nonrenormalization theorem. 
If the conjectured relation between 
Yang-Mills theory and supergravity is valid, 
it must be possible to derive the 
same property on the side of D3-brane Yang-Mills 
theory.  However, the special conformal 
transformation of the world-volume Yang-Mills 
theory is the standard one, 
\begin{eqnarray}
\delta_K x^a &=&  -2\epsilon\cdot x\, x^a +
\epsilon^a x^2  
\end{eqnarray}
without the last term of eq. (\ref{specialadsx}).  
On the Yang-Mills side, 
the coordinate $U$ is the radial component of the 
diagonal part of the 6 Higgs fields $X_i \, \, (i=1\sim 6)$ 
which, on the side of supergravity,  correspond 
to the space described by $\{U, S^5\}$. 

The solution of this puzzle is the following.  To study 
the dynamics of a probe D3-brane in the background of the 
source D3-branes, we have to derive effective 
theory for the diagonal Higgs fields by 
integrating over the off-diagonal components 
corresponding to the elements of the quotient group 
U($N$)/U($N-1$)$\times$U(1).   
In performing this integration, we have to impose the 
gauge condition, most conveniently, the familiar 
background gauge condition.  However, it turns out 
that the background gauge condition (or any other 
reasonable gauge condition) is not invariant 
under the special conformal transformation, and 
therefore we have to make a field-dependent 
gauge transformation which compensates 
the violation of the conformal invariance.  
Thus the transformation law of the diagonal 
Higgs fields receives a correction in a 
field-dependent manner.     In the large $U$ approximation, 
we can easily evaluate the correction by performing 
a one-loop calculation.  The final result 
precisely takes the form (\ref{specialadsx}) including the 
numerical coefficient.  For details, we 
refer the reader to ref. \cite{jky}. 
That the correct transformation law is obtained 
in the one-loop approximation including the 
precise coefficient suggests that some sort of 
the non-renormalization theorem is at work 
here, demanding that the 
lowest order result for the metamorphosed transformation law 
on the Yang-Mills side 
gets no higher oder corrections at least in the large $N$ limit.  

Since the derivation of the isometry almost amounts 
to the derivation of the background metric of the 
AdS$_5$,  the above result provides a strong support to the 
conjecture on the general relation 
between supergravity and 
supersymmetric Yang-Mills theory.  
In particular, the corrected transformation law 
explains the appeareance of the natural scale 
$(g_sN)^{1/4}$ from the side of Yang-Mills theory. 
If this result is not corrected by the higher order effect, 
it would be the first derivation of the 
scale which goes as $N^{1/4}$ in the large $N$ limit from a 
purely Yang-Mills point of view. 

This result also provides further evidence on our 
view about the relation between the source and the probe.  
Namely, the probe D3-brane is at somewhere far 
away from the horizon, while the AdS space itself is 
produced by the large number (=$N-1$) of the 
source D-branes at somewhere near (or inside ?) the horizon.  
  From this viewpoint, the operators corresponding 
to the boundary value of the bulk field 
for $U\rightarrow \infty$ are not, at least directly, 
the operators of the world-volume theory which 
corresponds to the large $N$ Yang-Mills theory.  
This raises a puzzle mentioned in the beginning of 
this subsection : Why and how does the boundary 
condition for large $U$ dictate the 
operators of the large $N$ Yang-Mills theory 
which corresponds to the source of the 
AdS space-time?  In the following, we suggest 
a simple argument which justifies the 
prescription of \cite{gkp}\cite{witten} 
under the assumption that the Maldacena's 
conjecture is true.  

The breaking of the gauge group U($N$) into 
U($N-M$)$\times$U(M) \,  ($N\gg M$) 
by assigning the large 
vacuum expectation value for the Higgs field 
corresponding to the radial direction amounts 
to introducing a heavy source and a light probe 
at a distance scale $U$ in the energy unit.  
We assume that the position of probe is at 
somewhere outside the near horizon limit, 
$U>R_c/\alpha'$.  
On the supergravity side, 
in the limit of large $N$ with fixed $M$, we can treat the 
effect of the probe as a small perturbation around 
the background of the heavy source. We thus 
decompose the metric as 
\EQ
g_{\mu\nu} = \overline{g}_{\mu\nu} + h_{\mu\nu}
\EN
where the first term $\overline{g}_{\mu\nu}$ 
is the classical metric produced by the 
source D3-branes and $h_{\mu\nu}$ 
is the metric produced by the probe in the 
background $\overline{g}_{\mu\nu}(N)$. The perturbative metric 
$h_{\mu\nu}$ satisfies the linearized equation in the lowest order 
approximation.  
\EQ
{\cal D}_N h_{\mu\nu}(u) = 2\kappa_{10}^2 T^p_{\mu\nu}(u)
\EN 
where $T^p_{\mu\nu}$ is the energy-momentum tensor 
of the probe 
\EQ
T^p_{\mu\nu}(u, x)
\propto {1\over \sqrt{-\overline{g}}}\delta^{(6)} (u-U)T^p_{\mu\nu}(x)
\EN
and $ {\cal D}_N$ is the kinetic operator 
for the linearized theory in the background of the 
source D3-branes.  We denote by $u$ 
the variable corresponding to the radial transverse 
coordinate in the bulk, while the 
coordinate along the D3-branes is denoted 
by $x$.  For notational simplicity, 
we suppress the angle variable corresponding to 
$S^5$. 
For the perturbative metric polalized along the direction parallel 
to the world volume, the kinetic operator 
essentially takes the following form 
\EQ
-{\cal D}_N = 
(1+2 g_{{\rm YM}}^2N\alpha'^2/r^4)^{1/2}\triangle^{\parallel}+
(1+2g_{{\rm YM}}^2N\alpha'^2/r^4)^{-1/2}\triangle_{R^6}
 \,  ,\EN
where $\triangle^{\parallel}$ is the 
laplacian for the flat four dimensions along the 
world volume and $\triangle_{R^6}$ is the 
flat six dimensional laplacian 
corresponding to the six dimensional transverse space. 
Note that the laplacian for the transverse part 
is proportional to the flat space laplacian as 
noted in \cite{douglas-taylor} even before taking 
the near horizon limit.  
The boundary condition for 
the linearized field (in the Euclidean 
formulation
\footnote{For a Lorentzian 
formulation, see 
\cite{bla}. }) is that it vanishes \cite{gkp} as $u\rightarrow 0$, 
since otherwise the solution diverges at the origin. 
By assuming that the states of the probe D3-brane can be 
chosen arbitrarily, the perturbative metric can also be 
assumed to induce an arbitrary boundary value  
$f_{\mu\nu}(x)$ at some 
large value of $u$.  It is natural to 
set the boundary at $u=R_c/\alpha'$ where the near-horizon 
limit loses its validity :   
\EQ
h_{\mu\nu}(u, x)\rightarrow 0 \, \, \, (u\rightarrow 0) \, \quad 
h_{\mu\nu}(u, x)\rightarrow  f_{\mu\nu}(x)\, \, \, (u\rightarrow R_c/\sqrt{\alpha'}).  
\EN
In the low-energy limit along the direction 
of the world-volume, we neglect the laplacian 
along the D3-brane (`quasi static' approximation) and we can approximate the 
boundary value as 
\EQ
f_{\mu\nu}(x) \propto \kappa_{10}^2{T^p_{\mu\nu}(x)\over (u-U)^4}\Bigr|_{u=R_c/\alpha'} \,  ,
\EN
since the laplacian for the transverse part of six dimensions is 
proportional to the flat space laplacian even outside the 
near horizon limit.   
The effective action for the boundary value is obtained 
by substituting the perturbed metric into the supergravity action, 
\EQ
S_{{\rm eff}}^{{\rm sg}}=
S[\overline{g}+h]  \, , 
\label{sugraaction}
\EN
using the AdS metric for the background. $\overline{g}_{\mu\nu}$. 
What we have done is essentially to replace the effect of 
the probe in arbitrary given states at somewhere 
outside the near horizon region by the boundary condition 
$f_{\mu\nu}$ for the perturbation $h_{\mu\nu}$ 
around the background of the source at 
the  boundary of the AdS space-time.  

Now, on the Yang-Mills side, we construct
 the effective theory for 
the unbroken part U($N-M$)$\times$U(M) after
 integrating over the heavy Higgs and 
W-bosons corresponding to the 
off-diagonal matrix elements 
whose `mass' is of order $U$ for large $U$ corresponding to 
the broken part of the gauge group.  
 This leads, again in the 
low-energy limit,  to 
the effective action 
\EQ
S^{{\rm YM}}_{{\rm eff}} \sim \kappa_{10}^2\int dx^4 {T^s_{\mu\nu}(x) T^p_{\mu\nu}(x)\over (u'-U)^4}
\EN
where $T^s_{\mu\nu} \sim \Tr_{N-M}
(F_{\mu\sigma}F_{\nu\sigma}) $ is the energy-momentum 
tensor of the source D3-branes on the Yang-Mills side. 
This one-loop result is exact because of the 
non-renormalization theorem \cite{dinesei} and 
hence is valid even for the probe at somewhere outside the 
near horizon region.  Note that this is consistent with 
the fact that the laplacian is proportional to that of 
the flat space in supergravity.   
Here we have assumed that the distance between the 
source and probe is  sufficiently large and is 
order of $|u'-U|$.  It seems natural to 
assume that 
$u'$ is of the same order as $R_c/\alpha'$.  
The source D3-branes cannot be considered to be at 
rest at the origin.  They are expected to extend to the 
whole range of the near-horizon region.  Then the 
average position of the source D3-branes would be determined 
by the scale $R_c$ which is the only scale in this region.   
Apart from a proportinal factor,  
we can replace the above expression by 
\EQ
S^{{\rm YM}}_{{\rm eff}} \sim \int dx^4  \, f_{\mu\nu}(x)T^s_{\mu\nu}(x)  \, .
\label{ymaction}
\EN
The equivalence between (\ref{sugraaction}) and 
(\ref{ymaction}), 
$
e^{-S_{{\rm eff}}^{sg}[f]} \propto 
\langle e^{-S^{{\rm YM}}_{{\rm eff}}[f] }\rangle \,   , 
$
 is essentially the statement of 
the usual prescription \cite{gkp} \cite{witten}. We have only discussed the metric perturbation, but the general idea can be easily extended to other massless fields.  Details remain to be 
seen. 

Our discussion clearly shows that the correlators 
we compute using the prescription \cite{gkp} \cite{witten} 
are those of the unbroken part U($N-M$) corresponding 
to the source, in spite of the fact that we use 
boundary conditions at the boundary of the 
AdS space-time.  It is not correct to think
 that the large $N$ Yang-Mills system is literally on the 
`boundary'.  
In order to derive the correlators from the boundary, 
we need in general 
definite rules which allow us to connect the 
operator insertions at the probe and 
the source.  
This is somewhat analogous to the LSZ relation 
between 
S-matrix elements and the corresponding 
Green functions.\footnote{
During writing the present manuscript, several 
works which are related to this issue 
appeared \cite{polchinski}\cite{susskind-holo}. 
Note, however, that the 
context of these recent works is slightly different from ours. } 
In our argument above, 
 the heavy `Higgs and W bosons' play the role of 
a `mediator' for connecting the probe and the 
source.   A similar reasoning also justifies the method for 
computing the Wilson loop expectation value,  proposed  in 
\cite{maldacena2} which naturally treats heavy 
W-bosons as `quarks' by utilizing the breaking U($N$)$\rightarrow$ 
U($N-1$)$\times$U(1) of gauge group as in our argument.   
In this case, the fundamental open string corresponding to 
the infinitely heavy W-boson, treated as a heavy 
point-like test particle, is playing 
the role of mediator.  

Our argument is based on the quasi-static approximation. 
This is justified for sufficiently large distances 
between the source and the probe, since the mass scale 
in terms of the world volume theory is very large and 
hence the characteristic distance scale with respect to 
the world volume  is small.  This is a manifestation 
of the space-time uncertainty relation 
explained in the next subsection.  
The correction to the quasi-static approximation can 
also be interpreted on the basis of the uncertainty relation, 
as discussed in \cite{douglas-taylor}: An uncertainty 
in the momenta along the world-volume is 
proportional to the uncertainty with respect to the 
transverse positions of the probe. 
Including this effect,   
more precise understanding on the correlators and also the 
extension to general D-branes are 
very important, since they may 
provide otherwise scarce information 
on the physics of the matrix models 
in the large $N$ limit.  For example, 
it would be extremely interesting if we 
can obtain some useful information 
on the large $N$ behavior of Matrix theory in this way
\cite{sek-yo}.  
In our argument, it is very crucial that the 
lowest order interaction between the source and the 
probe is equivalently described by both supergravity and 
matrix model even outside the near horizon region. 
For the validity of this property, the supersymmetric 
nonrenormalization theorem is important on the matrix side, 
while the laplacian must be essentially proportional to the 
flat space laplacian on the supergravity side. 
 It is not difficult to see that, from the viewpoint of 10 dimensions,  
the latter is satisfied for D-particles after taking 
into account the nontrivial behavior of dilaton.  
From the viewpoint of 11 dimensions, we have already 
seen this in the previous subsection.   

\subsection{Generalized conformal symmetry and 
space-time uncertainty principle}  

Next let us consider the question whether the 
conformal symmetry which is so important in the 
AdS/SYM relation has any generality beyond the 
special case of D3-branes.  One of the characteristics 
of Yang-Mills theories interpreted as the 
dynamical theory of D-branes is of course that the 
fields on the world-volume now represent the 
collective motion of D-branes in the bulk 
space-time.  This in particular implies that 
 the dimensionalities of the fields 
on the world-volume and of the base-space 
coordinates are opposite, as is seen from the 
transformation law (\ref{specialadsx}) and (\ref{specialadsu}), or more simply from the scale transformation
\EQ
X_i(x_a) \rightarrow X_i'(x_a') =\lambda X_i(x_a) ,
\label{eq21}
\EN
\EQ
x_a \rightarrow x_a' = \lambda^{-1} x_a . 
\label{eq22}
\EN
As is emphasized in ref. \cite{jy}, 
this indicates a general qualitative property 
that the long-distance phenomena in the (transverse) target space   is dual to the short distance phenomena in the world volume and 
{\it vice versa}.  
This property has also been emphasized independently
(named as 'UV-IR correspondence') 
 from the context of establishing the holographic 
bound for the entropy using the AdS/SYM 
correspondence in \cite{suss-witten}.  For a recent 
discussion on holography \cite{thooft-suss}, we refer the reader to \cite{susskind-holo}. 

Qualitatively, such a dual correspondence between the 
two different distance scales is precisely the prediction 
of the `space-time uncertainty principle' \cite{yostu}\cite{liyo}  
which has been 
proposed long ago as a possible 
space-time interpretation of the 
world-sheet conformal symmetry 
of perturbative string theory.  As already reviewed 
in some previous publications \cite{yo-1}\cite{liyo2},  
to which I would like to refer the reader  
for the explanation of the original motivation 
and examples,  the statement can be 
summarized as follows: 

\vspace{0.3cm}
\noindent
Let 
\begin{enumerate}
\item $\Delta T$ :  uncertainty in probing the distance scales in the \underline{longitudinal} directions along the world volumes of 
D-branes including time.

If the world-volume coordinates of D-branes 
in the static gauge 
are denoted by $x_a \, (a=0, \ldots, p)$, 
\[
\Delta T = |\Delta x| \]
 where 
$| \, \cdot \, |$ is the length in the  Euclidean metric. 
If we use the Minkowski metric, the original derivation of the 
relation requires that $\Delta T$ should be measured along the 
time-like direction along the world volume. 
 
\item $\Delta X$ : uncertainty in probing the distance scales 
in the bulk along the \underline{transverse} directions orthogonal to  
D-branes. 
\end{enumerate}
Then the following uncertainty relation is universally valid, 
\EQ
\Delta T \Delta X > \alpha'  .
\label{spacetimeuncertaintyrelation}
\EN

Note that this relation survives in the Maldacena limit 
($\Delta T \Delta U>1$), since 
\[
\Delta X \sim |\Delta r| = \alpha' |\Delta U|  .
\]
This explains the dual relation between 
the two different length scales determined by the mass 
of the open strings stretched between D-branes, on one 
hand, and the transverse distance between the branes, on the 
other. 
As discussed in \cite{douglas-taylor} and \cite{liyo2}, 
this elementary property is responsible for 
explaining  some important qualitative 
aspects of D-brane dynamics 
in connection with the AdS/CFT(SYM) correspondence and 
holography. 
Furthermore, 
if this relation is applied to D-particles, we can 
immediately derive the characteristic 
Planck scale $\ell_P=g_s^{1/3}\ell_s$ of 11 dimensions given only 
that the mass of a D-particle is of order $1/g_s\ell_s$ 
by combining with the ordinary quantum mechanical 
uncertainty relations.
This also leads to the holographic property that the 
minimum bit of information of the quantum state 
of a D-particle is stored in a cell of the order of 
the Planck volume in the 
transverse space in 11 dimensions.  

In particular, as is suggested in \cite{liyo2}, the space-time 
uncertainty relation can be regarded as an underlying principle 
behind the ultraviolet-infrared relation \cite{suss-witten} which is 
on the basis of the AdS/CFT(SYM) correspondence. 
Since the previous account 
given in \cite{liyo2} was ambiguous in some point, I would like to 
repeat the discussion here very briefly by taking into account the important 
observation made in \cite{peet-pol}. Our discussion on the 
correspondence between AdS$_5\times$S$^5$ and the 
type IIB string theory suggests that the uncertainty of the 
positions of the D3-branes in the radial direction $U$ is of order 
the AdS radius $R_c\sim (g_{{\rm YM}}^2N)^{1/4}\sim \Delta X$ in the string unit $\ell_s =1$. The space-time uncertainty relation then 
demands that the uncertainty with respect to the time-like 
length scale along the world volume is of order 
$\Delta T \sim 1/R_c$. Thus as the AdS radius increases, 
the dynamics of the D3-banes are probing 
 the high-energy region of the order 
$R_c$ of the AdS space-time. 
This is due to the fact that the typical mass scale of the 
open strings mediating the source D3-branes are growing as 
$R_c$. 
Does this imply that the length scale along the 
space-like directions on the world volume also decreases? 
Naively it might look so if we assume Lorentz invariance 
on the world volume.  However, the AdS/CFT relation 
leads to a contrary conclusion. From the viewpoint of 
the D-brane dynamics, the energy of the open 
strings mediating the interaction among the 
D3-branes can also be regarded as the self-energy of the 
heavy charged fields (U(1)). Let the uncertainty of the spatial 
position of such a charged field be $\Delta X_s$. 
The self energy of the field in the large $N$ strong 
coupling region can be estimated from the behavior of the 
Wilson loop \cite{maldacena2}, which tells us that it is of order 
$R_c^2/\Delta X_s$. Note that this is different from the 
weak coupling behavior which would be proportional to 
$R_c^4$: Coulomb force is still there corresponding 
to conformal symmetry but with the different effective 
charge.  Equating this result with 
the energy scale determined by the 
space-time uncertainty relation, we have 
\EQ
\Delta X_s\sim R_c.
\label{spacelikeuncertainty}
\EN
 Thus the world volume of the source D3-branes 
can be regarded as the collection of cells with volume $R_c^3$ 
in the space-like directions with a continuous flow of time, 
and hence the degrees of freedom of the theory 
is given, in terms of the 10 dimensional Newton constant 
$G_{10}\sim g_s^2\sim g_{{\rm YM}}^4$,  as 
$$
N_{dof}\sim N^2{L^3\over R_c^3}={L^3R_c^5\over G_{10}}
$$
where we have assumed that the source 
D3-branes wrap around a 3-torus of 
the length $L$ and also that 
the degrees of freedom is proportional 
to $N^2$ even in the strong coupling regime 
in view of the result \cite{gubklebtsyet} for the entropy at finite 
temperature. The final result is consistent with the Beckenstein-Hawking 
formula and is equivalent with the original result derived in \cite{suss-witten}. The relation (\ref{spacelikeuncertainty}) is 
at first sight quite surprising, but is an essential property 
ensuring holography. The above argument is consistent with the 
recent analysis \cite{polchinski}\cite{susskind-holo} 
of holography in the flat space limit. 
It should however be kept in mind that the space-time uncertainty 
relation or ultraviolet-infrared relation alone is 
not sufficient to derive holography. We have to 
combine it with some dynamical information, as  
exemplified by the assumptions needed in the 
argument. 

Although the space-time uncertainly 
relation might look at first sight too simple in order to characterize the 
short-distance space-time structure, 
it indeed captures the most important characteristics of quantum string theory including D-branes.  
We hope that it plays some role 
as one of the guiding principles toward 
nonperturbative formulation of string/M theory.  

The conformal symmetry can be regarded as a 
mathematical structure which characterizes the space-time 
uncertainty relation 
(\ref{spacetimeuncertaintyrelation}) 
: Clearly, the relation is invariant 
under the scale transformations 
$
\Delta T \rightarrow \lambda \Delta T , \quad 
\Delta X \rightarrow \lambda^{-1} \Delta X . 
$
The invariance can be extended to full conformal symmetry 
for general D$p$-branes, if we identify the 
uncertainty with the infinitesimal variations 
of the coordinate and fields, as 
\EQ
\delta_K \Delta T = -2\epsilon\cdot x \Delta T , 
\quad 
\delta_K \Delta X = 2\epsilon \cdot x \Delta X
\label{variationconf}
\EN
using the relation 
\[
\Delta (x_a +\delta_K x_a) =
-2\epsilon\cdot x \Delta x_a 
+2(\epsilon_a x\cdot dx - x_a \epsilon\cdot \Delta x)\, ,
\] 
where the second term is orthogonal to the first term 
and therefore we have the first equality of 
eq. (\ref{variationconf}).  
Thus it seems that the conformal symmetry 
plays an analogous role in the target space-time 
as that of the canonical structure in the phase space of 
classical mechanics.  This strongly suggests that 
the noncommutative nature of the space-time 
coordinates which characterizes the matrix 
models should be understood as a realization 
of the quantization of space-time and 
the conformal symmetry is a signature of certain 
unknown symmetry structure behind it.  
 
These considerations motivate us to generalize the 
conformal symmetry of D3-brane to general D-branes.  
Let us first consider the D-particle model. 
The action (\ref{d0matrixmodel}) is invariant under the 
scale transformation
\EQ
X_i(t) \rightarrow X_i'(t') =\lambda X_i(t) , \quad t\rightarrow t'=\lambda^{-1}t
\label{eq24}
\EN
\EQ
g_s \rightarrow g_s'=\lambda^3 g_s .
\label{eq25}
\EN
One might wonder whether this can be regarded as 
symmetry since we transformed the coupling constant 
simultaneously.  But this is not unreasonable if we remember 
that the string coupling constant, being given by the vacuum 
expectation value of dilaton at infinity, is not really a constant supplied by hand.  Ultimately the string coupling should 
be eliminated from the theory.  
From the viewpoint of 11 dimensions, the string coupling is 
replaced by the compactification radius, and the 
scale transformation can be 
reinterpreted as the boost transformation along the 
11th direction as follows.  In the above scale transformation, 
we have assumed that the string length is invariant.  
However, from the point of view of M-theory, we should fix the 
11 dimensional Planck length instead of the string length.  
This is achieved by redefining the unit of length 
as 
\[
\ell_s \rightarrow \lambda^{-1}\ell_s,\quad 
t\rightarrow \lambda^{-1}, \quad X_i\rightarrow 
\lambda^{-1} X_i, \quad A\rightarrow \lambda^{-1}A \, 
\]
where $A$ is  the gauge U($N$) gauge 
field.
Combining the change of unit, which does not change the action, with the above scaling transformation, the net transformation becomes 
\EQ
t\rightarrow \lambda^{-2} t, \quad 
R\rightarrow \lambda^2 R,\quad 
\EN
while the transverse coordinates and gauge field 
are scalar.  This is precisely the boost transformation 
provided we identify the time as the light-cone time $x^+$ 
and the compactification radius as that along the 
light-like direction $x^-$.  
  From this 11 dimensional viewpoint, it is 
more appropriate to express the space-time 
uncertainty relation in the form
$
R\Delta T \Delta X > \ell_P^3  
$
which suggests some `tripod'-like interpretation,  
possibly connected with the membrane structure, 
of the relation as already emphasized in \cite{liyo2}. 

Once we allow the 
variation of the string coupling, we can easily extend the 
symmetry to SO(2,1) group by considering the 
trivial time translation and the `special conformal' 
transformation whose infinitesimal form is 
\EQ
\delta_K X_i = 2  t X_i , \, \, \delta_K A= 2  t A , \, \, 
 \delta_K t =-  t^2 , \, \, \delta_K g_s =6  t g_s \,  .
\label{eq29}
\EN
   In all these transformations, the fermionic variables 
are assumed to be scalar.  

The above transformation property of the string coupling 
is essentially equivalent to the fact that the 
characteristic spatial and temporal 
scale of the dynamics of D-particle is proportional to 
$g_s^{1/3}\ell_s$ and $g_s^{-1/3}\ell_s$, respectively. 
Of course the inverse powers with respect to $g_s$ 
in these length scales just reflects the space-time 
uncertainty relation.   In contrast with this, 
there is no fixed characteristic scale in the case of D3-brane,  
because the dynamics is conformal invariant and 
all scales are equally important with respect to 
both $\Delta T$ and $\Delta X$. Of course, if we 
assume some particular background, the dynamics around the 
background can have characteristic scales.  
The scales which appeared in the case of 
D3-branes should be interpreted as such. 
We emphasize that this dual nature of two different 
scales in time and space explains the simultaneous 
appearance of the short distance scale $\Delta X$ and 
small energy scale $\Delta E\sim 1/\Delta T$ in the weak coupling dynamics of D-particles, ensuring the 
decoupling of the higher string modes in the short distance 
regime contrary to the naive intuition.

The argument discussed in the previous section 
connecting the D-brane Yang-Mills theory and 
supergravity should equally be valid for D-particles.  
Then we expect that the conformal symmetry 
of D-particle Yang-Mills theory must be 
reflected in the metric produced by a heavy 
source of D-particles.  The 10 dimensional 
metric around the 
D-particle is given, in the Maldacena limit 
$\alpha'\rightarrow 0, U=r/\alpha'$, by 
\EQ
ds_{10}^2 =\alpha'\Bigl(-
{U^{7/2}\over \sqrt{Q}}dt^2
+{\sqrt{Q}\over U^{7/2}}
\bigl(dU^2 + U^2 d\Omega_8^2\bigr)\Bigr)  ,
\label{eq38}
\EN
where   
\EQ
Q=60\pi^3 (\alpha')^{-3/2} g_sN=240\pi^5 g_{YM}^2 N .
\EN
In the case of D-particle, the dilaton is not constant and 
is given as
\EQ
e^{\phi} = g_s \Bigl({q \over \alpha'^7 U^7})^{3/4} 
=g_{YM}^2 \Bigl({Q\over U^7}\Bigr)^{3/4} . 
\label{eq39}
\EN
Both the metric and the dilaton are invariant 
under the dilatation
\EQ
U\rightarrow \lambda U ,\quad
 t \rightarrow \lambda^{-1}t ,\quad
 g_s \rightarrow \lambda^3 g_s .
\label{eq314}
\EN
Furthermore, they are also invariant under the 
infinitesimal special conformal 
transformation
\EQ
\delta_K t = -  \epsilon (t^2 +{g_{YM}^2N\over 96\pi^5 U^5}) , 
\label{eq315}
\EN
\EQ
\delta_K U =2 \epsilon tU ,\quad 
 \delta_K g_s=6 \epsilon t g_s  \, . 
\label{eq317}
\EN
Just as in the Yang-Mills case, these transformations 
with the time translation form an SO(2,1) algebra. 
The additional term $g_{{\rm YM}}^2N/ 96\pi^5 U^5$ 
in the special conformal transformation plays the 
similar role as in the case of D3-brane: 
The nonlinear field dependence is equally powerful 
to determine the effective action 
of the probe D-particle in the background of 
source D-particles.  We can  derive this 
modification of the transformation law 
in the bulk,  extending the similar mechanism as 
we have discussed for D3-brane in the previous 
subsection. For details about this, 
we refer the reader to 
refs. \cite{jy}\cite{jky2}.  It is straightforward to 
extend the conformal transformations 
of the above type to general Dp-branes (
$0\le p\le 4$), as discussed in the second of the latter references.  
The case of D-instanton matrix model, the so-called 
type IIB model \cite{ikkt}, is very special in this respect, since 
here all of the space-time coordinates are treated 
as matrices.  For an interpretation of the model  from the 
point of view of the space-time uncertainty relation and 
conformal symmetry, we refer the reader to \cite{yo-schild}.  

Finally, one might wonder  what is 
the relation of the 
space-time uncertainty relation and the associated 
conformal symmetries to supersymmetry. 
We can perhpas say that 
the supersymmetry is necessary to ensure 
some of prerequisites for applying the principle. 
For example, to discuss the scattering of D-branes 
meaningfully, it is necessary that the clusters  
far apart from each other should be free 
except for the weak gravitational forces among them. 
If the supersymmetry is not there, 
the quantum zero-point energy induces the 
forces which do not decay at large distances.  

\section{Graviton condensation in type IIB matrix model} 

As the final topic of this report, I would like to present some 
preliminary considerations on the treatment 
of graviton condensation  
in matrix models.  We have already seen some 
evidence that supersymmetric Yang-Mills models 
indeed describe the gravitational interactions of D-branes to 
certain extent.  However, it is clear that we do not have 
definite general principles which might explain the 
emergence of gravity from Yang-Mills theory. 
  From the viewpoint of symmetry, the existence of 
$N=2$ supersymmetry in space-time in 10 dimensions 
is the strongest 
argument for the existence of supermultiplet containing 
graviton, since only massless representation of the 
maximal $N=2$ supersymmetry in 10 dimensions 
is indeed the supergravity multiplet.  However, it is difficult 
to decide the presence of massless particles  
only from the logical structure of the Yang-Mills theory.  
In other words, without making concrete computations of D-brane scattering, 
we cannot decide whether 
the $N=2$ global symmetry is really 
elevated to the consistent local symmetry ensuring the emergence of gravity in the 
long-distance regime.  Since in general 
we expect that the matrix models are only 
sensible after taking appropriate large $N$ limit 
and then various questions can only be 
answered by solving complicated dynamics, it is 
very important to establish the symmetries 
of the models as far as possible.  

Now after seeing some evidence for the emergence of 
gravity in the Yang-Mills matrix models, we should 
be able to identify  the local space-time 
supersymmetry directly within the models.     
The purpose of the following preliminary consideration 
is to start an initial discussion toward such a 
possibility taking the simplest toy example of 
the type IIB matrix model.  We hope that our discussion 
will be a useful starting point for exploring possible 
higher symmetry structure in matrix-model 
approaches to non-perturbative string/M theory 
from a more general viewpoint.  

In the case of usual perturbative string 
theory,  that the theory is indeed a dynamical 
theory of space-time geometry is reflected on the 
fact that we can deform an allowed space-time background 
by insertion of the vertex operator 
corresponding to physical graviton modes of 
strings.  Or, if we use the language of string 
field theory,  the change of background is 
compensated by an appropriate redefinition 
of the string field corresponding to a shift of its 
graviton component.  In particular, the 
general coordinate transformation is 
compensated by such a field redefinition.\footnote{
For an initial discussion of this phenomenon, see 
\cite{y-stringfield}.}  
That is how the string theory can be generally 
covariant and in principle be 
a background independent 
formulation even if the theory is formulated without 
introducing the space-time metric explicitly as an 
independent degree of freedom.   
In the case of general Yang-Mills matrix models, 
on the other hand,  we cannot identify 
graviton modes directly in the classical 
action of the model.  They only appear as a part of 
loop effect in the `t-channel'.  For this reason, 
they can neither be treated as 
ordinary bound states, in general.  
In the special case of Matrix theory, 
only the Kaluza-Klein mode with non-zero 
11th momentum can be directly 
treated, and the graviton with 
zero 11th momentum can only appear as 
the loop effect.  

Let us now concentrate on the  
case of matrix model of D-instantons.  
The model is already Lorentz invariant and 
thus we can immediately ask a question, ``How is the 
symmetry extended to 
general coordinate invariance?".
The action of the model is 
\EQ
S_N=\Tr_{N} \Bigl(\, 
{1\over 4g^2}[X^{\mu}, X^{\nu}]^2 + 
{1\over 2}\overline{\Psi}\Gamma^{\mu}
[X^{\mu}, \Psi]\, 
\Bigr) .
\EN
Of course, it is 
not obvious what we mean by the general coordinate 
transformation for the matrix variables $X^{\mu}, \Psi$.  
In the usual interpretation,  only the diagonal 
components of $X^{\mu}$ have the meaning of the 
space-time coordinates and the off-diagonal 
components are really fields corresponding to the 
lowest open string modes.  In general, the 
space-time coordinate and the fields of open strings 
can have different transformation property.  
However, at least for the general linear 
transformations GL(10, R) which is globally 
defined, it is natural to 
suppose that the transformation law 
is the standard one 
\EQ
X^{\mu} \rightarrow a^{\mu}_{\nu} X^{\nu}\,  ,
\EN
where $a^{\mu}_{\nu}$ are arbitrary 
10$\times$10 coefficients.  
The action is manifestly invariant 
under the subgroup SO(9,1), if the spinor 
matrix $\Psi$ transformed as usual.  Our question is 
then whether it can be made invariant under the 
transformations belonging to the remaining broken quotient group 
GL(10, R)/SO(9,1).  
Of course, the standard procedure is to introduce 
the metric (or viel-bein) degree of freedom which absorbs the 
noninvariant piece of the action.  But as the metric degrees 
of freedom is supposed to be contained in the loop effect, 
there must exist different way of compensating the 
transformation without introducing the metric explicitly.  

Now we will present briefly 
an argument showing 
\cite{yo-prep}
that this can be achieved by embedding a 
model with fixed $N$  into models with larger $N$.  
The idea is to add more instantons to the model with 
appropriate information on the `state' of the 
added instantons such that they effectively produce 
the metric insertion for the original action with lower $N$.  
If we perform the embedding for all $N$ recursively, 
we naturally expect that the set of all such models 
as a whole, which we denote as 
$\{\ldots, U(N), U(N+1), \cdots\}$, 
can in principle describe all possible backgrounds 
of the model.  In this way, it should ultimately be possible to 
reconstruct the model in a background independent fashion.  

Let us study the simplest embedding from $N$ to $N+1$.  
We will use the following notations for the embedded matrices. 
\EQ
X^{\mu}_{a,N+1}=\phi^{\mu}_a ,
\quad 
X^{\mu}_{N+1, a}=\overline{\phi}^{\mu}_a ,
\quad 
X^{\mu}_{N+1,N+1}=x^{\mu} ,
\EN
\EQ
\Psi_{a, N+1}=\theta_a ,
\quad
\overline{\Psi}_{N+1, a}=\overline{\theta}_a , 
\quad
\Psi_{N+1,N+1}=\psi .
\EN
Thus the $N\times N$ matrices $X^{\mu}_{N\times N}, 
\Psi_{N\times N}$ are embedded into the corresponding 
$(N+1)\times (N+1)$ matrices as 
\[
X^{\mu}_{N\times N}  \ni 
\pmatrix{ . \qquad . \qquad . & . \cr
                    . \qquad . \qquad . & . \cr
                    . \quad \, \, \, X^{\mu}\, \,  \quad . & |\phi^{\mu}\rangle \cr
                    .\qquad  . \qquad  .  & . \cr 
                    . \qquad . \qquad . & . \cr
                    .\quad  \, \, \langle\phi^{\mu}| \, \quad .  & x^{\mu}\cr}_{(N+1)\times (N+1)}  ,
\]
\[
\Psi_{N\times N}  \ni
\pmatrix{ . \qquad . \qquad . & . \cr
                    . \qquad . \qquad . & . \cr
                    . \quad\, \, \,  \, \, \Psi \quad \, \,  \, . & |\theta\rangle \cr
                    .\qquad  . \qquad  .  & . \cr 
                    . \qquad . \qquad . & . \cr
                    .\quad  \, \, \, \langle\theta| \, \, \, \quad .  & \psi\cr}_{(N+1)\times(N+1)} .
\]
Here we use Dirac's bra-ket notation for the 
vector part.  Since the information on the state of 
the added instanton is specified by using the 
$N$-th diagonal elements $x, \psi$ (collective 
coordinates of the added instanton), 
we can first integrate over the 
vector parts to derive the effective action for the 
insertion of graviton.  Then we can further integrate over the 
collective coordinates
 by inserting an appropriate function $\Phi(x, \psi)$ whose form is determined later.  Let us call the 
result of this $\Delta \Gamma_N(X, \Psi;\Phi)$.  
Then by combining with the original U($N$) model 
$\exp (S_N) \rightarrow \exp (S_N) +\Delta \Gamma_N(X, \Psi;\Phi) $, 
we can define the new partition function 
of the ($N, N+1$) system.  
\EQ
Z_N [\Phi] 
={\cal N}^{-1}\int d^{10N^2}X d^{16N^2}\Psi 
\,  \exp \Bigl(
S_N[X, \Psi] +\sum_i c_i \Delta Z_N[X,\Psi;\Phi_i]\Bigl)
\EN
to the first order in the strengths $\{c_i\}$
 of the insertion, where the sum is over 
all independent `wave functions' of added instanton. 
It would be more appropriate to 
regard the wave functions as the scalar products of 
two wave functions.    

In the one-loop approximation, we can show that the 
following special choice of $\Phi$, 
which is the simplest candidate corresponding to the 
degree of freedom of symmetric tensor, gives the infinitesimal 
(first order) deformation of the action $S_N$ 
which compensates the change of the 
action by the infinitesimal 
GL(10,R)/SO(9,1)  coordinate transformation 
$a_{\mu\nu}=\eta_{\mu\nu} + S_{\mu\nu}$ 
where $S_{\mu\nu}$ is an arbitrary infinitesimal symmetric 
tensor. 
\EQ
\Phi(x, \psi) 
\propto 
(\Gamma^{\mu\beta\gamma} \Gamma^0)_{ab}
(\Gamma^{\nu\beta\gamma} \Gamma^0)_{cd} \, 
S_{\mu\nu}
 {\partial \over \partial \psi_a}{\partial \over \partial \psi_b} 
 {\partial \over \partial \psi_c}{\partial \over \partial \psi_d}
\Phi_0 
\EN
with 
\EQ
\Phi_0 = \prod_{a=1}^{16} \psi_a  \, \,  ,  \quad 
(\int d^{16}\psi\,  \Phi_0=
\langle 0| 0\rangle =1) .
\EN
Namely, apart from the proportional constants, 
the first order 
deformations  of the bosonic and fermionic part of the 
action are, respectively, 
\[
S_{\mu\beta}\Tr_N\Bigl([X^{\mu}, X^{\nu}]
[X^{\mu}, X^{\beta}]\Bigr) , \quad
S^{\mu\alpha}
Tr_N\Bigl(
\overline{\Psi}\Gamma^{\mu}[X^{\alpha}, \Psi]\Bigr)   \, 
\]
which are nothing but the change of the bosonic and 
fermionic actions 
corresponding to the quotient group GL(10,R)/SO(9,1). 
The one-loop approximation is justified 
for the wave functions which are constant 
with respect to the space-time coordinates, 
since then the infrared region $x\rightarrow \infty$ is 
dominant in the integral over the collective 
coordinate of the added instanton implying the 
infinite mass limit in the propagator 
of the fluctuating fields:  The coefficients 
of the above deformation are proportional 
to an infrared-divergent integral $\int d^{10}x/x^{10}$, 
which is cancelled by choosing the normalization of the 
wave function. 

Our result suggests how to  describe the change of 
background using only the degrees of freedom of 
the model itself, if we treat all possible embeddings  
simultaneously.  In particular, we have seen that 
the model indeed has full GL(10, R) symmetry.  
Thus the metric degrees of freedom 
appearing as a loop effect can be regarded as the Goldstone boson associated with 
the spontaneously broken 
part GL(10,R)/SO(9,1) 
of the infinitesimal symmetry GL(10,R) of the 
recursively embedded model 
$\{\cdots, U(N), U(N+1),\cdots \}$.   
Together with the space-time supersymmetry, this 
explains how the model  can indeed 
be the theory of gravity. 

\section{Concluding remarks}
In principle, the formalism suggested in the last section 
should be extended to 
arbitrary changes of background and hence to 
the definition of the model for general curved space-times.  
For example, the `wave function' 
$$
(\Gamma^{\mu\beta\gamma} \Gamma^0)_{ab}H_{\mu\beta\gamma}
{\partial \over \partial \psi_a}{\partial \over \partial \psi_b} \Phi_0
$$
describes the infinitesimal condensation of 
the  antisymmetric tensor field $B_{\mu\nu}$ 
with constant field strength $H_{\mu\nu\gamma}$. 
Also,  it is not difficult to derive the 
susy transformation law corresponding to the 
shift of the background. 
But, technically, computations required for such a 
generalization become increasingly difficult.  
I feel that we need some entirely new 
framework for developing the idea in a 
tractable way. What we are pursuing amounts to 
investigating the condensation of Goldstone bosons using the 
configuration space formalism.  Something 
which can play the role of the field-theory like 
formalism must be a desired language, by which we 
can treat the matrix models with different sizes of 
matrices in a much more unified and dynamical manner. 
Only by using such a formalism,  we would be able to 
discuss the major questions related to  
the present approach, such as the proof of S-duality 
symmetry, the background 
independent formulation, and so on.  
If we symbolically 
represent the whole recursive series 
$\{\cdots, U(N), U(N+1),\cdots \}$ by ${\cal H}[\Phi]$ 
as the functional of all possible background 
fields $\Phi$, background independence of the theory 
amounts to something like 
\EQ
`` \, \, {\delta {\cal H}[\Phi] \over \delta \Phi} =0 \, \, "
\EN
which should simultaneously play the role of the field equation 
in a perturbative approximation.  
We also note that our idea is intimately related to 
that\footnote{
For example, such a possibility has been 
suggested in \cite{douglas-ren}.} 
of  large-$N$ renormalization group, 
in which we try to derive the 
equation of motion for the background by imposing 
the fixed point condition in the sense of the 
renormalization group with respect to $N$.  In the latter, it 
is not clear how to define the model in curved space-time, 
and also how to treat the zero modes in formulating the
renormalization group.  Unless we insert `wave functions' 
as we have done, the result of embedding would only lead to 
a null result.  
In the approach suggested here, it may, at least in principle,  
be possible to develop the procedure in a more 
constructive fashion.  

Finally, it may be worthwhile to mention possible 
connection
\footnote{
I would like to thank J. Polchinski for calling my attention 
to Witten's work \cite{witten-k} at the Nishinomiya-Yukawa 
symposium.  
After submitting the present report to hep-th archive, 
I learnt the work \cite{horava} of Ho\v{r}ava which also suggested 
a possible connection of the K-theory formulation to 
Matrix theory.  I would like to thank P. Ho\v{r}ava 
for pointing out his recent works to me. } of the present idea with the K-theory 
formulation \cite{witten-k} of bound states 
of brane-anti-brane systems, 
which has been discussed to describe the 
stable non-BPS states \cite{sen}.  We should generalize the above construction such that the formalism includes  
not only variable number of D-instantons but also of 
anti-D-instantons simultaneously.  Obviously, the 
system with a fixed number of both 
D-branes and anti-D-branes 
cannot be supersymmetric.  It would be extremely 
interesting if we could recover supersymmetry 
by some similar mechanism as we have suggested 
for recovering the full 
GL(10,R) symmetry beyond manifest Lorentz symmetry.  
Such must also be crucial for developing 
covariant or background-independent 
formulation of Matrix theory.  I hope to report some 
progress along this line in near future.

\section*{Acknowledgements}
During the preparation of
 this report, I have enjoyed stimulating 
conversations with M. Ikehara, A. Jevicki, 
Y. Kazama, Y. Okawa, Y. Sekino and W. Taylor.    
The present work  is supported in part 
by Grant-in-Aid for Scientific  Research (No. 09640337) 
and Grant-in-Aid for International Scientific Research 
(Joint Research, No. 10044061) from the Ministry of  Education, Science and Culture.




\begin{thebibliography}{99}
\bibitem{pol} J. Polchinski,  Phys. Rev. Lett. 75  
(1995) 4724.
\bibitem{wittensft} E. Witten, Nucl. Phys. B268 (1986) 253. 
\bibitem{bfss} T. Banks, W. Fischler, S. H. Shenker and L. Susskind, 
Phys. Rev. D55 (1997) 5112. 
\bibitem{dhn} B.  de Witt, J. Hoppe and H.Nicolai, 
Nucl. Phys. B305[FS23] (1988) 545. 
\bibitem{index} P. Yi, hep-th/9704098.\\
S. Sethi and M. Stern, hep-th/9705046. 
\bibitem{suss}  L. Susskind, hep-th/9704080.
\bibitem{seisen} N. Seiberg, Phys. Rev. Lett. 79(1997) 3577. \\
A. Sen, hep-th/9709220.
\bibitem{dkps} M. Douglas, D. Kabat, P. Pouliot and S. Shenker,
Nucl. Phys. B485(1997)85.
\bibitem{pss} S. Paban, S. Sethi and M. Stern, hep-th/9805018, hep-th/9806028. 
\bibitem{lowe} D. Lowe, hep-th/9810075. 
\bibitem{bbpt} K. Becker and M. Becker, Nucl. Phys. B506 (1997) 48 \\
K. Becker, M. Becker, J. Polchinski and A. Tseytlin, 
Phys. Rev. D56 (1997) 3174. 
\bibitem{oy1} Y. Okawa and T. Yoneya, Nucl. Phys. B538(1998) 67. \\
See also,  \\
M. Dine and A. Rajaraman, Phys. Lett. B425 (1998) 77. \\
 M. Fabbrichesi, G. Ferretti and R. Iengo,  J. High Energy Phys. 06 (1998) 002; hep-th/9806166.
\\
W. Taylor IV and M. V.  Raamsdonk, hep-th/9806066.\\
R. Echols and J. P. Gray, hep-th/9806109.
\\
J. McCarthy, L. Susskind and A. Wilkins, hep-th/9806136.
\bibitem{oy2} Y. Okawa and T. Yoneya, hep-th/9808188.
\bibitem{daniel} Ulf H. Danielsson, M. Kruczenski and P. Stjernberg, 
hep-th/9807071. 
\bibitem{dine} M. Dine, R. Echols and J. P. Gray, hep-th/9810021. 
\bibitem{douglas} For a review and further references, see 
M. Douglas, hep-th/9901146. 
\bibitem{taylor} W. Taylor IV and M. V. Raamsdonk, 
hep-th/9812239. 
\bibitem{dewit} B. de Wit, K.Peeters and J. Plefka, hep-th/9803209. 
\bibitem{maldacena} J. Maldacena, hep-th/9711200. 
\bibitem{itzak} N. Itzhaki, J. Maldacena, J. Sonnenshain and 
S. Yankielowicz, hep-th/9802042. 
\bibitem{gkp} S. S. Gubser, I. R. Klebanov and A. M. Polyakov, hep-th/9802109.
\bibitem{witten} E. Witten, hep-th/9802150.
\bibitem{jky} A. Jevicki, Y. Kazama and T. Yoneya, Phys. Rev. Lett. 81 (1998) 5072. 
\bibitem{maldacena2} J. Maldacena, hep-th/9803002. 
\bibitem{douglas-taylor}  M. Douglas and W. Taylor IV, hep-th/980725. \\
See also,\\
S. R. Das, hep-th/9901004.
\bibitem{bla} V. Balasubramanian, P. Kraus and 
A. Lawrence, hep-th/9805171. 
\bibitem{dinesei} M. Dine and N. Seiberg, hep-th/9705057. 
\bibitem{polchinski} J. Polchinski, hep-th/9901016.\\
V. Balasubramanian, S. B. Giddings and A. Lawrence, hep-th/9902052. 
\bibitem{susskind-holo} L. Susskind, hep-th/9901079.
 \bibitem{jy} A.Jevicki and T. Yoneya, Nucl. Phys. B535 (1998) 335. 
\bibitem{suss-witten} L. Susskind and E. Witten, hep-th/9805114.
\bibitem{thooft-suss} G. 'tHooft,  gr-qc/9310026.\\
L. Susskind, J. Math. Phys.  36 (1995)6377.
\bibitem{sek-yo} Y. Sekino and T. Yoneya, work in progress. 
\bibitem{yostu} T. Yoneya, {\it Duality and Indeterminacy Principle in String 
Theory}
in ``Wandering in the Fields", eds. K. Kawarabayashi and
A. Ukawa (World Scientific, 1987), p. 419: see also
{\it String Theory and Quantum
Gravity} in ``Quantum String Theory", eds. N. Kawamoto
and T. Kugo (Springer, 1988), p. 23; Mod. Phys. Lett.  A4, 1587(1989).
\bibitem{yo-1} T. Yoneya, hep-th/9707002 (to appear in the
Proceedings of the APCPT-ICTP Joint International Conference '97);
\bibitem{liyo} M. Li and T. Yoneya,   Phys. Rev. Lett. 78
 (1997) 1219 
\bibitem{liyo2} M. Li and T. Yoneya,  hep-th/9806240. 
\bibitem{peet-pol} A. W. Peet and J. Polchinski, hep-th/9809022. 
\bibitem{gubklebtsyet} S. S. Gubser, I. Klebanov and A. A. Tseytlin, 
hep-th/9805156. 
\bibitem{jky2} A. Jevicki, Y. Kazama and T. Yoneya, hep-th/98010146. \\
See also a recent calculation which went 
beyond the eikonal approximation,\\
H. Hata and S. Moriyama, hep-th/9901034. 
\bibitem{ikkt} N. Ishibashi, H. Kawai, Y. Kitazawa and A. Tsuchiya,
Nucl. Phys B498(1997)467.
\bibitem{yo-schild} T. Yoneya, Prog. Theor. Phys. 97(1997) 949.
\bibitem{y-stringfield} T. Yoneya, Phys. Rev. Lett. 55(1985) 1828; 
Phys. Lett. 197B(1987) 76. 
\bibitem{yo-prep} T. Yoneya, in preparation. 
\bibitem{douglas-ren} See, e.g., M. Douglas, hep-th/9707228. 
\bibitem{witten-k} E. Witten, hep-th/9810188. 
\bibitem{horava} P. Ho\v{r}ava, hep-th/9812135, and a talk at 
APS meeting, UCLA, Jan., 1999. 
\bibitem{sen} A. Sen, hep-th/9805019; hep-th/9808141. \\
O. Bergman and M. R. Gaberdiel, hep-th/9806155. 



\end{thebibliography}
\end{document}